\documentclass[prd,aps,twocolumn,nofootinbib,superscriptaddress,eqsecnum,floatfix,preprintnumbers,amsmath,amssymb,
nofootinbib,longbibliography,]{revtex4-1}

\usepackage{amssymb,amsmath}
\usepackage{epsfig}
\usepackage[dvipsnames]{xcolor}
\usepackage[utf8]{inputenc}
\usepackage{stmaryrd}
\usepackage{mathrsfs}
\usepackage{mathalfa}
\usepackage{accents}
\usepackage[normalem]{ulem}

\usepackage{graphicx}
\graphicspath{ {c:/Documents/PhysicsNotes/GradResearch/Images/GR_Lens/} }

\newcommand{\pa}{\partial}

\newcommand{\be}{\begin{equation}}
\newcommand{\ee}{\end{equation}}
\newcommand{\bea}{\begin{eqnarray}}
\newcommand{\eea}{\end{eqnarray}}
\newcommand{\ba}{\begin{equation}\begin{aligned}}
\newcommand{\ea}{\end{aligned}\end{equation}}

\newcommand{\beg}{\begin{gather*}}
\newcommand{\eng}{\end{gather*}}
\newcommand{\hh}{,\hspace{0.5cm}}
\newcommand{\hhh}{,\hspace{0.2cm}}

\newcommand{\lap}{\triangle}
\newcommand{\lapp}{{}^{(2)\!}\triangle}
\newcommand{\n}[1]{\label{#1}}

\newcommand{\const}{\mbox{const}}

\newcommand{\ts}[1]{{\boldsymbol{#1}}}
\newcommand{\ie}{\emph{i.e.} }

\def\XXint#1#2#3{{\setbox0=\hbox{$#1{#2#3}{\int}$ }
\vcenter{\hbox{$#2#3$ }}\kern-.6\wd0}}

\usepackage[makeroom]{cancel}
\usepackage[caption=false]{subfig}
\usepackage[colorlinks=true,
            citecolor=green,
            linkcolor=red,
            filecolor=cyan,
            urlcolor=magenta,
            backref=false]{hyperref}

\newcommand{\GG}[3]{\Gamma_{#1#2}^{\ \ #3}}

\begin{document}

\title{Gravitational lensing, memory and the Penrose limit}

\author{Valeri P. Frolov}
\email{vfrolov@ualberta.ca}
\affiliation{Theoretical Physics Institute, University of Alberta, Edmonton, Alberta, Canada T6G 2E1}
\author{Alex Koek}
\email{koek@ualberta.ca}
\affiliation{Theoretical Physics Institute, University of Alberta, Edmonton, Alberta, Canada T6G 2E1}

\date{\today}

\begin{abstract}
In this paper, we discuss the gravitational field of ultrarelativistic extended spinning objects. For this purpose, we use a solution of the linearized gravitational equations obtained in the frame where such an object is translationally at rest, and boost this solution close to the speed of light. In order to obtain a regular limiting metric for non-spinning matter, it is sufficient to keep the energy of the boosted body fixed. This process is known as the Penrose limit. We demonstrate that in the presence of rotation, an additional rescaling is required for the angular momentum density components in the directions orthogonal to the boost. As a result of the Lorentz contraction, the thickness of the body in the direction of the boost shrinks. The body takes the form of a pancake, and its gravitational field is localized in the null plane. We discuss light and particle scattering in this gravitational field, and calculate the scattering parameters associated with the gravitational memory effect. We also show that by taking the inverse of the Penrose transform, one can use the obtained scattering map to study the gravitational lensing effect in the rest frame of a massive spinning object.

\medskip

\hfill {\scriptsize Alberta Thy 4-22}
\end{abstract}

\maketitle

\section{Introduction}

In this paper we discuss properties of the gravitational field of ultrarelativistic spinning objects. For this purpose, we first describe the field of a massive spinning object in the frame where it is at rest.
We do not assume that its rotation is rigid, and allow the matter within the object to have differential rotation as well. We use the weak field approximation so that the corresponding gravitational field can be found as a solution to the set of linear partial differential equations in the flat background metric. This material is well known, and can be found in many textbooks on General Relativity (see e.g. \cite{Misner:1974qy,Poisson:2004,Carroll:2003,FrolovZelnikov:2011}).

In order to find the gravitational field of an ultrarelativistic spinning compact object, one can proceed as follows: First, one performs a boost transformation; that is, one makes a Lorentz transformation to a moving inertial reference frame. In the limit where the velocity parameter tends to the speed of light, the obtained boosted metric becomes singular. However, for a non-spinning body, this limiting metric can be made regular if in the process of boosting one keeps fixed not the mass of the object, but its energy. This special procedure is known as a Penrose limit. In a more general setup, this procedure was considered in detail by Penrose \cite{Penrose1976}.

In the simplest case, when a boosted body is spherically symmetric and its gravitational field is decribed by the Schwarzschild metric, the obtained limiting metric coincides with the Aichelburg-Sexl metric \cite{Aichelburg:1970dh}. This solution can be also obtained as a gravitational shock wave due to a massless particle moving at the speed of light \cite{Dray:1985} (see also \cite{Voronov,Barrabes:2001}).

There exists many publications which are devoted to the generalization of these results for boosted Kerr and Kerr-Newman black holes. In the first of these papers \cite{Lousto:1989,Ferrari1990,Lousto1992}, the simplest case was studied where the boost direction coincides with the angular momentum of the black hole. In addition, the rotation parameter was taken fixed in the Penrose limit. As a result, the angular momentum of the black hole $J=Ma$ vanished in this limit, and the resulting metric described the gravitational field of massless particles located at a "ring singularity" of radius $a$. Different approaches and discussions of the boosted Kerr black holes can be found in \cite{,Hayashi:1994,Barrabes:2003,Barrabes:2004,Soares:2019,Gallo:2020}.

It should be emphasized that some controversy in the results of different approaches to the above problem is partly connected with the following general problem.
There always exists an ambiguity connected with possible rescaling of the parameters of the original metric. Namely, when one considers the limit of a family of
solutions to Einstein equations as some free parameter of these solutions approaches a certain value, one can always make a parameters-dependent coordinate transformation before taking this limit. As a result, one can arrive at different limiting metrics. This problem is discussed in detail in a nice paper \cite{Geroch:1969}.

In addition to the limiting boosted black hole metric, solutions for extended one dimensional ultrarelativistic objects have been constructed and studied.  A  solution for an ultrarelativistic object which has finite length in the direction of motion, \ie a light beam or ``pencil", was obtained by Bonnor \cite{Bonnor1969}. To obtain this metric, instead of solving the Einstein equations, one can boost the metric of a line object. However, in this process one should rescale its length parameter as well in order to compensate for the Lorentz contraction of the pencil. Let us note that a pencil is the simplest version of an extended body, which is extended in one direction and has vanishing size in the transverse directions. This idealization greatly simplifies calculations.

The gravitational field of a spinning pencil of light was obtained by Bonnor in 1970 \cite{Bonnor:1970sb}, see also \cite{doi:10.1063/1.528380,PhysRevD.96.104053}. Higher-dimensional solutions describing the gravitational field of spinning ultrarelativistic objects and light beams were found in \cite{Frolov:2005in,Frolov:2005zq}. In these papers such spinning ultrarelativistic objects received the name  ``gyratons'',  which is now often used in the literature.
Let us emphasize that in these type of solutions, one imposes an additional assumption. Namely, it is assumed that the angular momentum of the object is directed along its velocity. Such a component of the angular momentum is not transformed under the adopted boost, and so no additional rescaling of the metric parameters is required in the Penrose limit (see e.g. \cite{FrolovZelnikov:2011,Boos:2020}.)

In the present paper we generalize these results in two ways. We study the Penrose limit of the metrics for extended objects with differential rotation of its matter, and we do not assume that this rotation occurs only in two-planes orthogonal to the direction of motion. In this case, the components of the angular momentum density orthogonal to the direction of motion do not vanish. We demonstrate that during the process of taking the Penrose limit, one needs to make an additional rescaling of the parameters connected with these transverse components of the angular momentum.

As a result of the Lorentz contraction, the boosted body takes the form of a pancake, and the original information about its matter density and its rotation becomes encoded in the 2D surface characteristics of the squeezed pancake. The gravitational field of such an object is also squeezed and located within the 3D null plane, which we denote by $\Gamma$. To find this field we solve the corresponding gravity equations,
and after taking the proper Penrose limit, we obtain the gravitational field of such an ultrarelativistic object. This is one of the results presented in this paper.

The second subject which is considered in this paper, is scattering of null rays and particles in the obtained metric. Null ray and particle worldlines are straight lines before and after they meet the null plane $\Gamma$. We demonstrate that while passing through this plane, two effects occur: the rays and particles change their direction of motion, and there exists an additional shift of the position of the outgoing worldline just at the moment when it passes through $\Gamma$.
These effects are manifestations of the well known gravitational memory effect.
When we compare the distances between freely moving objects as well as their relative velocities
before the passage of a gravitational wave burst versus after, we find that there is a difference  \cite{Thorne:1992sdb}. This effect, called the gravitational memory effect, has been widely discussed both in application to gravitational wave astronomy \cite{Favata:2010zu,Pasterski:2015tva,Zhang:2017rno,Zhang:2017geq}, and in the general framework of gravitational field theory \cite{Strominger:2017zoo,Hollands:2016oma}. The memory effect exists not only for pure gravitational waves but also for the gravitational field of ultrarelativistic objects and gyratons \cite{Shore:2018kmt}. In this paper, we demonstrate that for an ultrarelativistic extended and spinning source, the change in position and velocitiy of passing particles and null rays does not only depend on the on mass and angular momentum of the source. It also depends on the matter \emph{density} and angular momentum \emph{distribution} within the source.
We also discuss the memory effect for the case where the impact parameter of the incoming ray is much larger than the transverse size of the boosted body, which greatly simplifies our description.

Finally, we demonstrate that the obtained results for the scattering of null rays by the gravitational field of an ultrarelativistic spinning object can be used to study the scattering of light by a massive spinning object in its own reference frame.
For this purpose, one can use the obtained results for null rays propagating in the gravitational field of the ultrarelativistic extended spinning object, and perform an inverse Penrose transformation. This transformation consists of the inverse boost transformation, which is accompanied by a proper rescaling of the matter and angular momentum densities.  This observation can be used for studying the gravitational lensing effect \cite{Bartelmann:1999yn,LENS1,Bozza:2002zj,Perlick:2004tq,LENS2,Bozza_2010,Bartelmann:2016}.
In particular, we demonstrate that in the far-distance approximation, the positional shift and angular deflection from the scattering coincide with the results obtained for the scattering of light in Kerr spacetime.

This paper is organized as follows. The gravitational field of a massive spinning object in the linearized gravity approximation is discussed in Sections~\ref{2} and \ref{3}. Boosting the extended objects and the Penrose limit are considered in Sections~\ref{4}--\ref{6}. Particle and light scattering, and the gravitational memory effect are discussed in Section~\ref{7}. In Section~\ref{8}, we discuss how the obtained results for light scattered by ultrarelativistic objects can be used to study the gravitational lensing effect in the weak field of a compact massive spinning object.
In Section~\ref{9}, we discuss a large distance approximation for the scattered light. Section~\ref{10} is devoted to discussion of the obtained results. In appendix~\ref{A}, we construct asymptotic expansions of scattering angles and shift parameters for null rays with large impact factor in the Kerr metric.  This is done for both equatorial and azimuthal rays. Appendix~\ref{B} collects useful formulas and contains a derivation for the ray equation of motion in the geometry studied in this paper.

In this paper we use sign conventions adopted in \cite{Misner:1974qy}. We also use units in which $G=c=1$. We restore Newton's coupling constant $G$ in some final expressions.

\section{Gravitational field of massive spinning objects in linearized gravity}
\n{2}

Let us consider a flat spacetime, and write its metric in the form
\be
ds_0^2=\eta_{\mu\nu}dX^{\mu} dX^{\nu}=-dT^2+dX^2 +dY^2+dZ^2\, .
\ee
We suppose that there is a stationary compact massive object, and denote by $T_{\mu\nu}$ the stress-energy tensor of its matter distribution. We denote by $\ts{\xi}=\partial_T$ a timelike Killing vector. Then for a stationary matter distribution, one has
\be \n{STAT}
{\cal L}_{\xi}\ts{T}=0\, ,
\ee
where ${\cal L}_{\xi}$ is the Lie derivative along vector $\ts{\xi}$.

In the presence of matter, the spacetime metric takes the form
\be
g_{\mu\nu}=\eta_{\mu\nu}+h_{\mu\nu}\, ,
\ee
where $h_{\mu\nu}$ is a perturbation of the metric induced by the matter distribution.
It is convenient to introduce new variables $\bar{h}_{\mu\nu}$ for the metric perturbations
\be
\bar{h}_{\mu\nu}={h}_{\mu\nu}-{1\over 2}\eta_{\mu\nu} {h}_{\lambda}^{\lambda}\, ,
\ee
and impose the de Donder gauge fixing condition
\be
\bar{h}^{\mu\nu}_{\ \ ,\nu}=0\, .
\ee
Then the linearized Einstein equations take the form
\be \n{BOX}
\Box \bar{h}_{\mu\nu}=-16\pi T_{\mu\nu}\, .
\ee
These equations can also be rewritten as follows
\be \n{hhTT}
\Box {h}_{\mu\nu}=-16\pi \bar{T}_{\mu\nu}\hh
\bar{T}_{\mu\nu}={T}_{\mu\nu}-{1\over 2}\eta_{\mu\nu} {T}_{\lambda}^{\lambda}\, .
\ee

For the stationary distribution of matter, that is when condition (\ref{STAT}) is satisfied, the metric perturbation does not depend on time and equation (\ref{BOX}) reduces to the following relation
\be
\lap {h}_{\mu\nu}=-16\pi \bar{T}_{\mu\nu}\, .
\ee

Since the background metric $\ts{\eta}$ is flat, it therefore is Poincare invariant, and has 10 Killing vectors that allow one to introduce the following conserved quantities
\ba
&P^{\mu}=\int T^{0\mu} \, d^3 X\, ,\\
& J^{\mu\nu}=\int (X^{\mu}T^{\nu 0}-X^{\nu}T^{\mu 0} )\, d^3 X\, .
\ea
We choose our reference frame so that the system as a whole is at rest, and so one has
\be \n{PPP}
P^{i}=\int T^{0 i} \, d^3 X=0\hh i=1,2,3\, .
\ee
Then the total mass $M$ of the system is
\be \n{MMM}
M=\int T^{0 0} \, d^3 X\, .
\ee
We also choose the origin of the spatial coordinate to be at the center-of-mass of the system, so that the following condition is satisfied
\be
\int X^k T^{00} \, d^3 X=0\, .
\ee
Here and later we use small Latin letters $i, j, k$ for spatial indices which take values $1, 2, 3$.

For this choice of the coordinates, one has
\be \n{JJJ}
J^{kl}=-J^{lk}=2\int X^k T^{l0} \, d^3 X\, .
\ee
This antisymmetric 3D tensor is related to the angular momentum vector $\vec{L}$ of the system as follows
\be
L_k={1\over 2} e_{klm} J^{lm}\, ,
\ee
where $ e_{klm}$ is a totally antisymmetric 3D Levi-Civita tensor.  A diagram of the matter distribution in its rest frame is shown in figure \ref{Penrose1} below.

\begin{figure}[!hbt]
    \centering
      \includegraphics[width=0.4\textwidth]{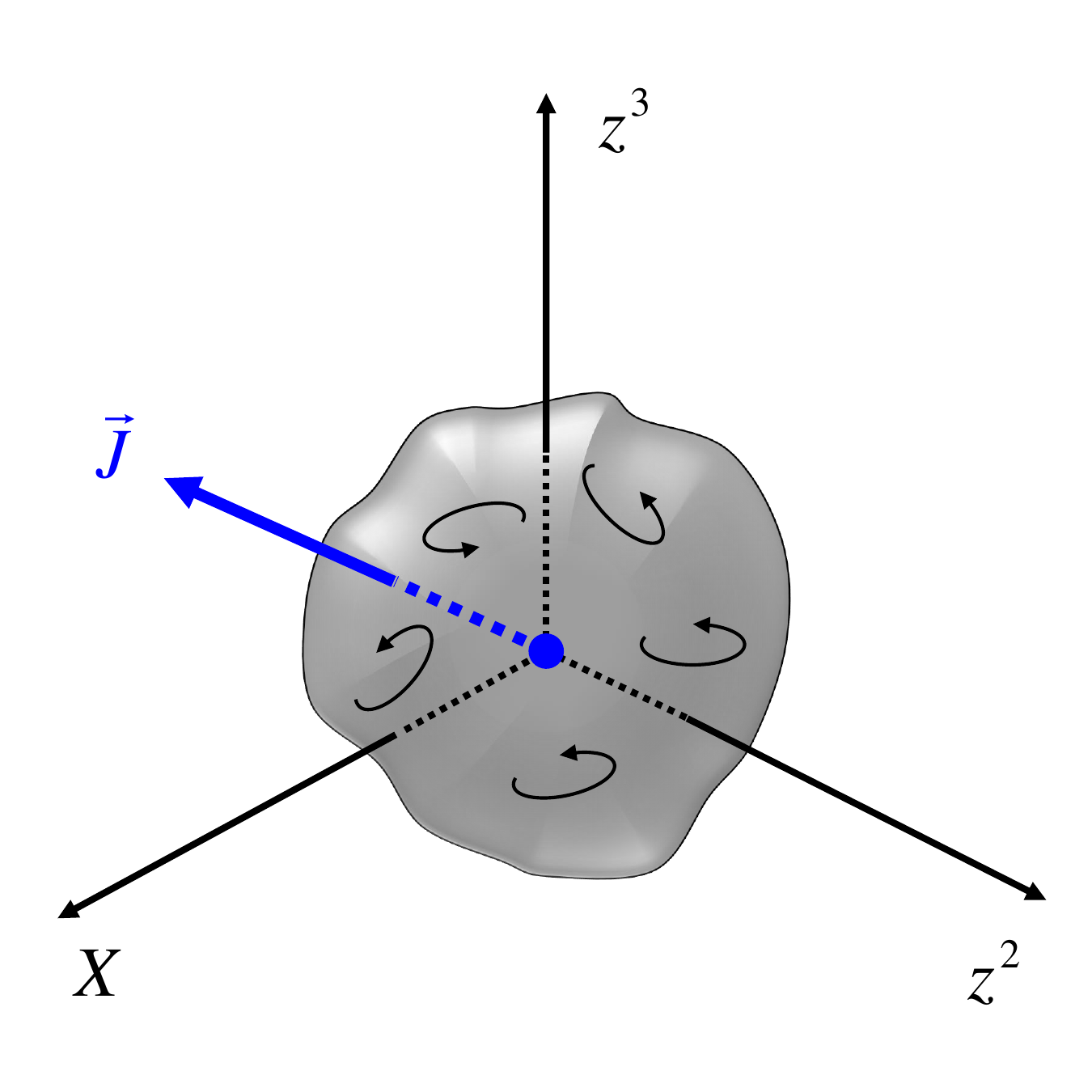}
    \caption{Extended massive spinning object.}
    \label{Penrose1}
\end{figure}

Let us denote by $L$ the size of the compact object. Then it is easy to show that at the distance $r\gg L$, the metric perturbation is of the form
\be
h_{00}\approx {2M\over r}\hhh h_{ik}\approx {2M\over r}\delta_{ik}\hhh  h_{0k}\approx {2J_{kl}X^l\over r^3}, .
\ee
By a rigid spatial rotation, the antisymmetric matrix $J_{kl}$ can be transformed into the following canonical form
\be
J_{kl} =
\begin{pmatrix}
0 & J &  0 \\
-J&0 &0 \\
0 & 0 & 0
\end{pmatrix}\, .
\ee
Using spherical coordinates one can write the asymptotic form of the metric as follows
\ba
&ds^2\approx -\left(1-{2M \over r}\right) dT^2-{4J \sin^2\theta\over r} dT d\phi \\
&+\left(1+{2M \over r}\right)(dr^2+r^2 d\omega^2)\, ,
\ea
where $d\omega^2$ is the standard metric for the unit 2-sphere.

\section{Stress-energy tensor of a compact massive body with differential rotation}

\n{3}

We use the following ansatz for the stress-energy tensor
\be \n{TENN}
T_{\mu\nu}=\xi_{(\mu} K_{\nu)}\, ,
\ee
where
\be
K_{\nu}=\rho \xi_{\nu}+{\cal Z}_{\nu}\hh \xi^{\nu}{\cal Z}_{\nu}=0\, .
\ee
Since the vector $\ts{{\cal Z}}$ is orthogonal to the Killing vector $\ts{\xi}$, it has the following form ${\cal Z}_{\nu}=(0,\vec{{\cal Z}})$. We choose the 3D vector $\vec{{\cal Z}}$ as follows
\be
{\cal Z}_i=j_{ik}^{\ \ ,k}\, .
\ee
Both the scalar function $\rho$ and the antisymmetric tensor $j_{ik}$ are time-independent. It is easy to check that the stress-energy tensor (\ref{TENN}) satisfies the conservation law
\be
T^{\mu\nu}_{\ \ ,\nu}=0\, .
\ee

The non-vanishing components of the stress-energy tensor are
\be
T_{00}=\rho\hh T_{0i}={1\over 2} {\cal Z}_i\, .
\ee
Equation (\ref{MMM}) shows that $\rho$ signifies the mass density, while $j_{ik}$ is associated with the angular momentum distribution. To demonstrate this, one can use the relations
\be
J^{kl}=\int X^k {\cal Z}^l d^3 X=\oint_{S} X^k j^{lm} d\sigma_m -\int j^{lk} d^3 X\, .
\ee
Taking the boundary surface $S$ to be outside the body, one can set the surface integral equal to zero, and therefore
\be
J^{kl}=\int j^{kl} d^3 X\, .
\ee

 One also has
 \be
 \int T^{0 i} \, d^3 X={1\over 2}\int {\cal Z}^i\, d^3 X={1\over 2}\oint_{S}  j^{im} d\sigma_m=0\, .
 \ee
This relation implies that (\ref{PPP}) is valid.

\section{Boosting the source}

\n{4}

To obtain the metric of an ultrarelativistic spinning object, we shall perform a boost transformation. Namely, we shall use two inertial reference frames which we denote  by $S'$ and $S$. The reference frame $S'$ is chosen so that the center-of-mass of the object is at rest at its origin ${O}$. We call  $S'$ a rest frame. The other frame, $S$,  moves with respect to the rest frame with a constant velocity.
We choose coordinates $(X,Y,Z)$ in $S'$ so that the velocity vector $\vec{V}$ of $S$ points along the negative $X$-axis, $\vec{V}=(-V,0,0)$.  We denote the Killing vector in the boost direction by $\ts{\zeta}=(0,1,0,0)$.

It is convenient to present 4D Minkowski spacetime as a direct sum of two 2D spaces.
The first 2D space $\Pi_1$ is spanned by Killing vectors $\ts{\xi}$ and $\ts{\zeta}$, while the other 2D space $\Pi_2$ is orthogonal to it. The boost transformation acts in $\Pi_1$, while $\Pi_2$ is not affected by the boost. In accordance with this, we project 4D vectors and tensors onto the 2D planes $\Pi_1$ and $\Pi_2$. We shall use the capital letters $A,B,\ldots$  for the indices $0,1$  associated with $\Pi_1$, while the lowercase letters $a,b,\ldots$ denote indices $2,3$ associated with $\Pi_2$.

We denote
\be
{\cal Z}_X=\zeta^{\mu}{\cal Z}_{\mu}\, ,
\ee
then one has
\be
{\cal Z}_{\mu}={\cal Z}_X\zeta_{\mu}+{\cal Z}_a \delta^a_{\mu}\, .
\ee
The antisymmetric tensor $\ts{j}$, which enters the expression (\ref{TENN}), allows a similar decomposition. If one of its indices takes the value 1, then the other index should be either 2 or 3. If both indices of $\ts{j}$ belong to the $\Pi_2$ plane, then it is proportional to the 2D Levi-Civita symbol $e_{ab}$. Following this observation, we write
\be
j_a=j_{\mu a}\zeta^{\mu}\hh j_{ab}=e_{ab}j\, .
\ee
Then one has
\be
{\cal Z}_X=j^a_{,a}\hh {\cal Z}_a=-\pa_X j_a+e_{ab}j^{,b}\, .
\ee

Using the above notations, the stress-energy tensor  (\ref{TENN}) can be written as follows
\be \n{TEN1}
T_{\mu\nu}=\rho \xi_{(\mu} \xi_{\nu)}+{\cal Z}_X  \xi_{(\mu} \zeta_{\nu)}+{\cal Z}_a \xi_{(\mu} \delta_{\nu)}^a\, .
\ee

Our next goal is to obtain the expression for the components of $T_{\mu\nu}$ in the boosted frame.
We denote by $(t,x)$ the coordinates in the $\Pi_1$ plane of the boosted frame $S$. Then one has
\be\n{LT}
X=\gamma(x-\beta t)\hhh T=\gamma(t-\beta x)\hhh Y=y\hhh Z=z\, .
\ee
Here, $\beta=V/c$ and $\gamma=(1-\beta^2)^{-1/2}$.
We shall also use the notation $z^a$, $a=2,3$, for the coordinates transverse to the boost direction.

Let us denote
\ba
&U={1\over \sqrt{2}}(T-X)\hh V={1\over \sqrt{2}}(T+X)\, ,\\
&u={1\over \sqrt{2}}(t-x)\hh v={1\over \sqrt{2}}(t+x)\, .
\ea
Then the Lorentz transformation (\ref{LT}) implies
\ba\n{UVTX}
&U=\alpha u\hh V=\alpha^{-1} v\hh \alpha=\sqrt{1+\beta\over 1-\beta}\, ,\\
&T={1\over \sqrt{2}}(\alpha u+\alpha^{-1}v)\, ,\  X={1\over \sqrt{2}}(-\alpha u+\alpha^{-1}v)\, .
\ea
For ultrarelativistic motion in the frame $S$, that is when $\beta\to 1$, one has
\be \n{aagg}
\alpha=2\gamma\left(1-{1\over 8}\gamma^{-2}+O(\gamma^{-4})\right)\, .
\ee

The trace of the stress-energy tensor is
\be
T_{\mu}^{\mu}=-\rho\, .
\ee
Hence, the tensor $\bar{T}_{\mu\nu}$ defined by (\ref{hhTT}) takes the form
\be
\bar{T}_{\mu\nu}={1\over 2} \rho Q_{\mu\nu}+{1\over 2} {\cal Z}_X P_{\mu\nu}+S_{\mu\nu}+N_{\mu\nu}\, .
\ee
Here
\ba
 &Q_{\mu\nu}=\xi_{\mu}\xi_{\nu}+\zeta_{\mu}\zeta_{\nu}\hh  P_{\mu\nu}=\xi_{\mu}\zeta_{\nu}+\zeta_{\mu}\xi_{\nu}\, ,\\
& S_{\mu\nu}={\cal Z}_a \xi_{(\mu} \delta_{\nu)}^a\hh N_{\mu\nu}={1\over 2}\rho \delta_{\mu}^a\delta_{\nu}^b\delta_{ab}\, .
\ea

Using the expressions for $\xi_A$ and $\zeta_A$
\be
\xi_A=(-1,0)\hh \zeta_A=(0,1)
\ee
one gets
\be
Q_{AB}=
\begin{pmatrix}
1 &  0 \\
0&1
\end{pmatrix}\hh
P_{AB}=
\begin{pmatrix}
0 & -1 \\
-1&0
\end{pmatrix}\, .
\ee

We use the notation $x^{\mu'}\!=(u,v,z^a)$ for the new boosted null coordinates. Since the boost does not affect the transverse directions, one has $z^a=X^a$. The components of $\ts{Q}$ and $\ts{P}$ in these new coordinates are
\be
Q_{A'B'}=
\begin{pmatrix}
\alpha^2 &  0 \\
0&\alpha^{-2}
\end{pmatrix}\hh
P_{A'B'}=
\begin{pmatrix}
\alpha^2 & 0 \\
0& -\alpha^{-2}
\end{pmatrix}\, .
\ee
We can transform the components of $\ts{S}$ from the stationary frame to the boosted frame in a similar fashion as above, and the components of $\ts{N}$ stay the same between the two frames. Thus
\ba
&\bar{T}_{uu}={1\over 2}(\rho+{\cal Z}_X)\alpha^2\hh
\bar{T}_{vv}={1\over 2}(\rho-{\cal Z}_X)\alpha^{-2}\, ,\\
&\bar{T}_{ua}=-{1\over 2\sqrt{2}}\alpha {\cal Z}_a\hh
\bar{T}_{va}=-{1\over 2\sqrt{2}}\alpha^{-1} {\cal Z}_a\, ,\\
&\bar{T}_{ab}={1\over 2}\rho \delta_{ab}\, .
\ea
The other components of $\bar{T}_{\mu'\nu'}$ vanish.

\section{Penrose limit of the stress-energy tensor}

\n{5}

\begin{figure}[!hbt]
    \centering
      \includegraphics[width=0.35\textwidth]{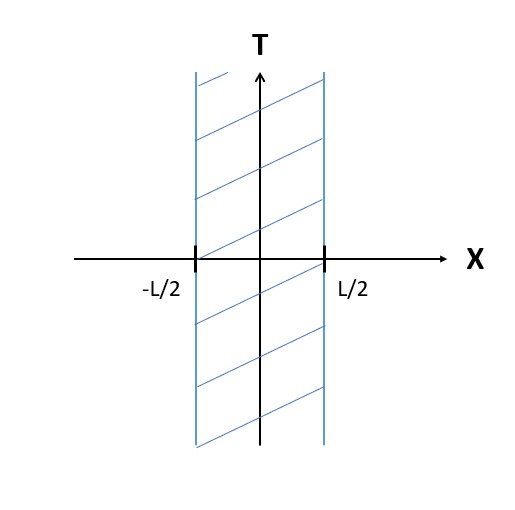}
    \caption{$TX$-diagram. Extended object has size $L$ in the $X$-direction in its rest frame.}
    \label{Penrose2}
\end{figure}

\begin{figure}[!hbt]
    \centering
      \includegraphics[width=0.35\textwidth]{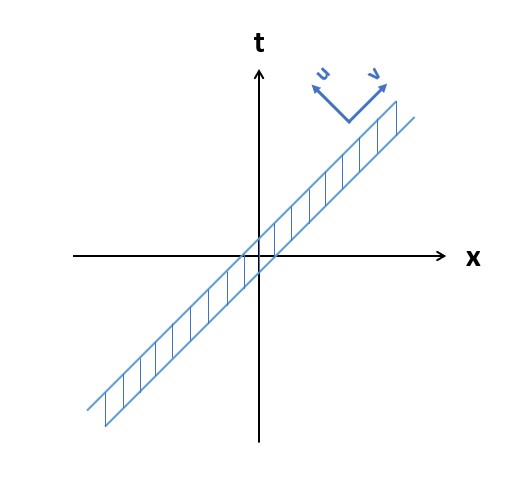}
    \caption{$tx$-diagram. The same extended object in the moving frame is squeezed in the direction of motion. }
    \label{Penrose3}
\end{figure}

We consider compact massive objects. This means that their matter is localized within a compact region. We denote by $L$ the characteristic size of this region, so that the stress-energy tensor in the $S'$ frame vanishes outside a sphere of radius $L$.  The size of this region is frame-dependent, as seen by comparing figures \ref{Penrose2} and \ref{Penrose3}.

\begin{figure}[!hbt]
    \centering
      \includegraphics[width=0.4\textwidth]{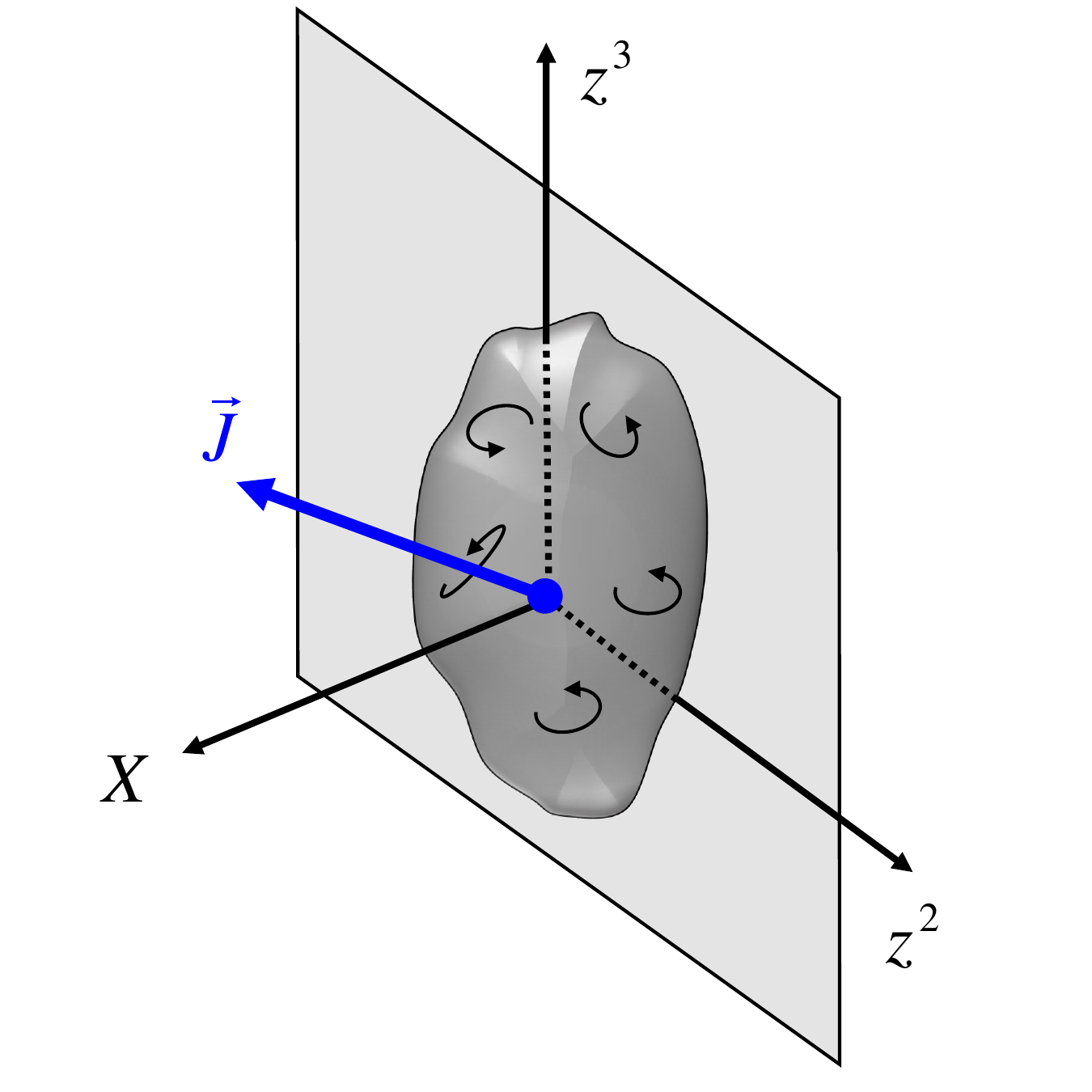}
    \caption{Boosted massive spinning object. As a result of Lorentz contraction, it takes the form of a 2D spinning pancake.}
    \label{Penrose4}
\end{figure}

As a result of the Lorentz contraction, the size of the body shrinks in $X$-direction, so that it looks like a pancake when observed from the $S$ frame (see figure \ref{Penrose4}). Let us discuss this effect in more detail.

Let us consider a function $f(X,z)$ \footnote{
Here and later we use $z$ as shorthand for $z=(z^2,z^3)$.
}, which vanishes outside the interval $X\in (-{L/2},{L/ 2})$, and let us denote by $I(z)$ the following integral taken along $v=0$ line
\be \n{intb}
I(z)=\int du \alpha f(X,z) B(u)\, .
\ee
Here $B(u)$ is a slowly changing function of the retarded time $u$, which near $u=0$ allows the following expansion
\be
B(u)=B(0)+\left. {dB\over du}\right|_{u=0} u+\ldots \, .
\ee
Using (\ref{UVTX}) for large $\alpha$, one gets\footnote{Here and later we denote by $\simeq$ an equality valid in the leading order of an $\alpha^{-1}$ expansion.}
\ba
&I(z)\simeq -\sqrt{2}{\cal F}(z) B(0)\, ,\\
& {\cal F}(z)=\int dX f(X,z)\, .
\ea
In other words, in the calculation of the integral (\ref{intb}) along  null rays performed in the leading order of the parameter $\alpha^{-1}$, one can use the following substitution
\be
\alpha f(X,z)\to - \sqrt{2} \delta(u)\ {\cal F}(z)\, .
\ee

In order to obtain a finite expression for the metric in the ultrarelativistic limit  $\gamma\to\infty$, one should perform a special rescaling of the source parameters. We denote
\be
\hat{\rho}={\alpha}\rho\hh \hat{j}_a={\alpha}{j}_a\, ,
\ee
and keep the functions $\hat{\rho}$ and $\hat{j}_a$ fixed. In the leading order in $\alpha$, one has
\be
\hat{M}=\int \hat{\rho}\ d^3X=\alpha M\, .
\ee
For ultrarelativistic motion, $\alpha\approx 2\gamma$, which means that $\hat{M}\approx 2E$, where $E$ is the energy of the boosted body.
Hence keeping $\hat{\rho}$ fixed implies that the energy of the boosted body $E$ is constant. This is nothing but the  assumption adopted in the definition of the Penrose limit.

Let us denote
\be
\hat{\mu}=\int (\hat{\rho}+\hat{{\cal Z}}_X) dX\hh
\lambda_a=e_{ab}\int j^{,b} dX\, .
\ee
Then the leading non-vanishing terms of the stress-energy tensor $\bar{\ts{T}}$ are
\be \n{TTPP}
\bar{T}_{uu}\simeq -{1\over\sqrt{2}}\delta(u)\, \hat{\mu}(z)\hh
\bar{T}_{ub}\simeq {1\over 2}\delta(u)\, \lambda_b(z)\, .
\ee

\section{Metric in the Penrose limit}

\n{6}

The background metric in $(u,v,z^a)$ coordinates is
\be
ds_0^2=-2 du\, dv+ dz_a dz^a\, .
\ee

\begin{itemize}
\item Since the stress-energy tensor (\ref{TTPP}) does not depend on the coordinate $v$, the perturbation $h_{\mu'\nu'} $ must have the same property, and equations (\ref{hhTT}) reduce to
\be
\lapp {h}_{\mu\nu}=-16\pi \bar{T}_{\mu\nu}\, .
\ee
Here $\lapp$ is a 2D transverse Laplacian
\be
\lapp={\pa^2\over \pa (z^2)^2}+{\pa^2\over \pa (z^3)^2}\, .
\ee
\item Since the only non-vanishing components
of the stress-energy tensor (\ref{TTPP}) in the leading order are
$\bar{T}_{uu}$ and $\bar{T}_{ua}$, the perturbed metric has the form
    \ba \n{METUV}
   & ds^2=-2 du\, dv+ dz_a dz^a\\
    &+\Phi(u,z) du^2+2A_a(u,z)du\, dz^a\, .
    \ea
The metric of this form and its higher-dimensional generalizations were studied in the papers \cite{Frolov:2005in,Frolov:2005zq}.
\item Since the components of stress-energy tensor (\ref{TTPP}) are  products of $\delta(u)$ and functions depending on $z^a$, one has
    \ba
   &\Phi(u,z)=\delta(u)F(z)\, ,\\
   &A_a(u,z) =\delta(u){\cal A}_a(z)\, .
    \ea
\end{itemize}

The metric coefficients $F(z)$ and ${\cal A}_a(z)$ obey equations
\ba
&\lapp F=8\sqrt{2}\pi  \hat{\mu}\, ,\\
&\lapp {\cal A}_a=-8\pi \lambda_a\, .
\ea
Solutions of these equations can be found by using the Green function $G(z,z')$ of the 2D Laplace operator
\be
\lapp G(z,z')=\delta^2(z-z')\, ,
\ee
This Green function is
\be
G(z,z')={1\over 4\pi}\ln (|z-z'|^2/C^2)\, ,
\ee
where $|z-z'|^2= |z^2-z'^2|^2+ |z^3-z'^3|^2$ and $C$ is an arbitrary constant.
Hence, one has
\ba\n{FFAAEQ}
&F(z)=2\sqrt{2}\int \ln (|z-z'|^2/C^2)\hat{\mu}(z') \,d^2z'\, ,\\
&{\cal A}_a(z)=-2\int  \ln (|z-z'|^2/C^2) \lambda_a(z')\, d^2z'\, .
\ea
\section{Particle and null ray scattering}

\n{7}

\subsection{Scattering problem}

\begin{figure}[!hbt]
    \centering
      \includegraphics[width=0.4\textwidth]{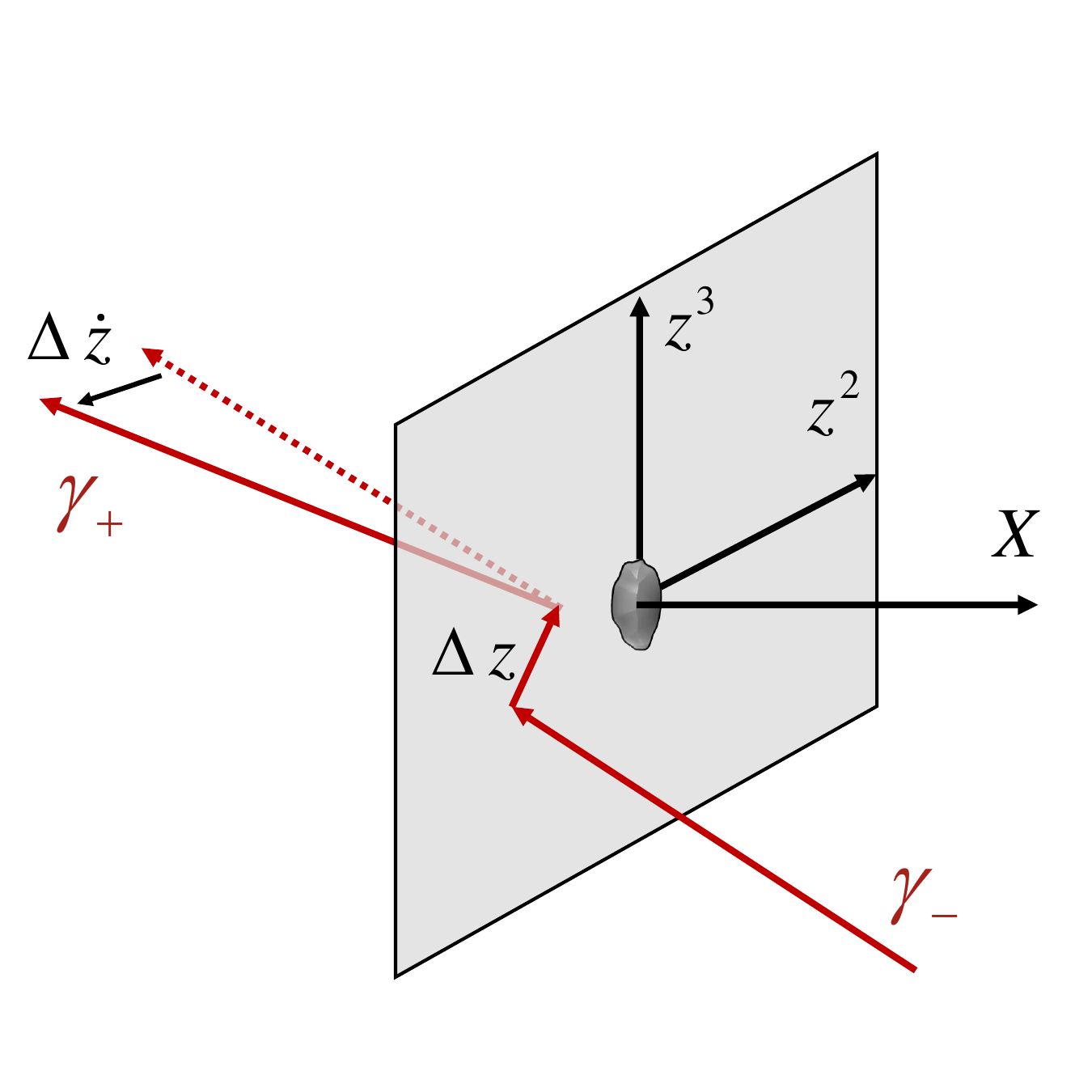}
    \caption{Null ray scattering}
    \label{Penrose5}
\end{figure}

The gravitational field of the ultrarelativistic spinning object is localized within the null plane $u=0$. We denote this 3D null surface by $\Gamma$. This surface $\Gamma$ separates two spacetime domains $M_{\pm}$, which we call past and future. Both of them are parts of Minkowski spacetime.
In $M_{\pm}$, free massive and massless particles move along straight-line geodesics. We denote by $\gamma_-$ a part of such a geodesic before the particle "meets" $\Gamma$, and by $\gamma_+$ its part after it passes through  $\Gamma$ (see figure \ref{Penrose5} below for more detail).

A null ray in flat spacetime can be characterized by 5 parameters. To specify a ray, it is sufficient at a given moment of time to give a spatial point through which it passes (3 parameters), and a direction of the ray at this point (2 parameters). For a massive particle, we need 6 parameters. In addition to direction of particle motion, one should specify the value of its speed.

We write a worldline of a particle in the form $x^{\mu}=(u(\tau),v(\tau), z^a(\tau))$,
where $\tau$ is an affine parameter along the particle's trajectory. Since the metric (\ref{METUV}) does not depend on $v$, and $\pa_v$ is its Killing vector, the quantity ${\cal U}=du/d\tau$ is an integral of motion.
For a massive particle, the parameter $\tau$ coincides with the proper time, and ${\cal U}$ is an additional nontrivial constant which characterizes the particle's motion.
For a null ray, there exists a rescaling ambiguity for the affine parameter $\tau$, and so one can always put $u=\tau$. In what follows, we shall use this parametrization.

 Geodesic equations in the  metric (\ref{METUV}) are discussed in Appendix~\ref{B}. To apply these results to our case, it is sufficient to put
\be \n{ppqq}
p(u)=q(u)=\delta(u)\, .
\ee
Then equations (\ref{VVV})  and (\ref{ZZZ}) of Appendix~\ref{B} imply
\ba   \n{ZZDD}
&\ddot{z}_a+\delta(u){\cal F} e_{ab}\dot{z}^b+\dot{\delta}(u){\cal A}_a-{1\over 2}\delta(u)F_{,a}=0\, ,\\
&2\dot{v}=\dot{z}_a \dot{z}^a+\delta(u)(F+2{\cal A}_a\dot{z}^a)+{\varepsilon\over {\cal U}^2}\, ,\\
& {\cal U}=du/d\tau=\mbox{const}\, .
\ea
The first of these relations is a set of two second-order ordinary differential equations for $z^a(u)$, while the second relation is a first order equation for $v=v(u)$. The set of equations (\ref{ZZDD}) uniquely specifies a solution after giving 5 quantities as initial conditions. For a null ray, this is a complete set of required parameters. For the case of a massive particle, one additionally needs to specify the constant ${\cal U}$.

Using retarded time $u$ as a parameter along the worldline, we write solutions of these equations in $M_{\pm}$ subspaces as follows
\ba\n{zzvvpm}
&z^a_{\pm}=c^a_{\pm}+\dot{c}^a_{\pm}\ u\, ,\\
&v_{\pm}=b_{\pm}+\dot{b}_{\pm}\ u\, .
\ea
Here $c^a_{\pm}$, $\dot{c}^a_{\pm}$, $b_{\pm}$ and $\dot{b}_{\pm}$ are constant parameters. The normalization equation, that is the second equation in (\ref{ZZDD}), implies that
\be
\dot{b}_{\pm}={1\over 2}\left( c^a_{\pm}c_{\pm a}+{\varepsilon\over {\cal U}^2}\right)\, .
\ee

To satisfy the equations (\ref{ZZDD}), one can search for $z^a(u)$ and $v(u)$  in the following forms
\ba \n{zzvv}
&z_+^a(u)=z^a_-(u)+\Delta c^a\theta(u)+\Delta \dot{c}^a u\,\theta(u)\, ,\\
&v_+(u)=v_-(u)+\Delta b \theta(u)\, .
\ea
Here $\theta(u)$ is a Heaviside step function.
Then one has
\ba
&\dot{z}_+^a(u)=\dot{z}_-^a(u)+\Delta c^a\delta(u)+\Delta \dot{c}^a \theta(u)\, ,\\
&\ddot{z}_+^a(u)=\ddot{z}_-^a(u)+\Delta c^a\dot{\delta}(u)+\Delta \dot{c}^a \delta(u)\, ,\\
&\dot{v}_+=\dot{v}_-(u)+\Delta b \ \delta(u)\, .
\ea
In the absence of the matter distribution on $\Gamma$, the quantities $\Delta c^a$, $\Delta \dot{c}^a$ and $\Delta b$ vanish. In the weak field approximation adopted in this paper, the quantities $F$ and ${\cal A}_a$ are assumed to be small. Similarly, the jump parameters $\Delta c^a$, $\Delta \dot{c}^a$ and $\Delta b$ are small as well. This means that in the leading order, the products of these quantities with $F$ and ${\cal A}_a$ can be neglected.
Substituting (\ref{zzvv}) into (\ref{ZZDD}) and keeping the leading order terms, one obtains
\ba\n{ANSW}
&\Delta \dot{c}^a={1\over 2}F^{,a}-{\cal F} e^{a}_{\ b} c_{-}^b\, ,\\
&\Delta c^a=-{\cal A}^a\hh \Delta b={1\over 2} F+{\cal A}_a\dot{c}^a_{-}\, .
\ea

The obtained results mean that in the general case, after passing through the $\Gamma$ plane, the parameters of the particle and light trajectories are changed:
\begin{itemize}
\item Their spatial position on $\Gamma$ is shifted by the value $\Delta c^a$;
\item Their direction of motion is changed. The parameter which controls this change is $\Delta \dot{c}^a$.
\item There exists also a shift $\Delta b$ of the coordinate $v$, which is related to the time-delay effect.
\end{itemize}

\subsection{Gravitational memory effect}

Past $M_-$ and future $M_+$ domains separated by a null plane $\Gamma$ are just two copies of a part of Minkowski spacetime. Particles and light rays freely propagating in $M_-$ are also free moving in $M_+$. However, their relative positions and velocities are different.  One can say that there exist a map between in and out states of the particles and rays, and this map depends on the properties of the ultrarelativistic object. This is a very special case of a general effect known as the gravitational memory effect. Similar situations happen when, instead of an ultrarelativistic particle, a burst of pure gravitational waves propagates in flat (or curved) spacetime \cite{Thorne:1992sdb}. This effect was widely discussed in the literature since it might have possible applications in the search for gravitational waves \cite{Favata:2010zu,Pasterski:2015tva,Zhang:2017rno,Zhang:2017geq}.
This and similar effects are also interesting from the point of view of general field theory, because they are connected to encoding information in the soft modes of the fields
\cite{Strominger:2017zoo,Hollands:2016oma}.

In the case discussed in the present paper, initial and final states of massive and massless particles are described by the independent constants which enter relations (\ref{zzvvpm})
\be
\psi_{\pm}=\{ c^a_{\pm},\dot{c}^a_{\pm},b_{\pm},({\cal U})\}\, .
\ee
Formulas (\ref{ANSW}) establish a relation between in and out states of the particles and light
\be \n{PSI}
\Psi: \psi_-\to\psi_+\, .
\ee
This map depends not only on "global" parameters such as the energy and angular momentum of the ultrarelativistic object, but also on the details of its internal structure, such as the energy density and spin distribution inside the body. It is quite an interesting question: Suppose one knows the corresponding map $\Psi$ for all possible configurations of particles and null rays.
How can detailed information about the structure and properties of the ultrarelativistic object be obtained?

\section{Gravitational lensing and the inverse Penrose transform}

\n{8}
To obtain the gravitational field of an ultrarelativistic object, we performed the Penrose transform. Namely, we used equations (\ref{UVTX}) to obtain the relation between the rest-frame coordinates $(T,X)$ and  boosted null coordinates $(u,v)$. At the same time, we rescaled the parameters of the mass density $\rho$, transverse $j^a$, and longitudinal $j$ spin density parameters
\be
{\cal P}:  (T,X)\to (u,v)\, ,\
\hat{\rho}=\alpha\rho\, ,\ \hat{j}^a=\alpha j^a\, ,\ \hat{j}=j\, .
\ee
When $\alpha\to\infty$, the map ${\cal P}$ describes the Penrose limit, provided the quantities $\hat{\rho}$,  $\hat{j}^a$ and $\hat{j}$ are fixed.

In section~\ref{7}, we obtained relations between in and out trajectories of light rays for their scattering in the gravitational field of an ultrarelativistic object. Let us demonstrate now how this map $\Psi$, (\ref{PSI}), can be used for the solution of another problem: the gravitational lensing of light by a compact massive spinning object. We show that for this purpose, it is sufficient to perform the inverse transformation of ${\cal P}$, which we call the inverse Penrose transform.

To illustrate this, let us consider a beam of light rays which move parallel to the $X$ axis before they meet the null plane $\Gamma$. For such rays, $z_-^a=c^a_-=$const. We call $\ell=\sqrt{c^a_- c_{a -}}$ the impact parameter. Using the results of section \ref{7}, one finds that after the incoming null ray passes through the plane $\Gamma$, so that it is in the domain $u>0$, its equation is
\be
z_+^a(u)=c_-^a+\Delta c^a+\Delta \dot{c}^a u\, .
\ee
Since $X\approx -{\alpha\over \sqrt{2}}u$, the scattering angles in the rest frame $S'$ are
\be\n{DDAA}
\Delta\hat{\theta}^a=-\lim_{u\to \infty}{z_+^a(u)\over X}={\sqrt{2}\over \alpha} \Delta \dot{c}^a\, .
\ee
For $\alpha\to \infty$ these scattering angles tend to zero. However, if one applies the inverse Penrose transform, one should restore the original rest frame quantities $\rho$ and $j^a$, and this "enhances" the angle $\Delta\hat{\theta}_a$ by multiplying it by $\alpha$. After this, the limit for the scattering angles $\Delta{\theta}_a$ in the rest frame becomes finite.
Performing the calculations, one obtains for the scattering angles in the rest frame $S'$ the following expression
\ba \n{INT}
&\left.\Delta\theta^a=2{\pa \over \pa z^a}H(z)\right|_{z^a=z_0^a}\, ,\\
&H=\int dX'd^2z' \ln (|z-z'|^2) \\
&\times\left[\rho(X',z')+\pa_{a'} j^{a'}(X',z') \right]\, .
\ea
We omit the constant factor $C^2$ inside the logarithm since its  contribution to $\Delta\theta^a$ vanishes.

Let us summarize.  One can use the derived expressions for the scattering of light by an ultrarelativistic object in order to obtain the scattering angles for  light rays propagating near a massive spinning object in its rest frame. This result gives an alternative option for studying the gravitational lensing effect in the weak field approximation \cite{Bartelmann:1999yn,LENS1,Bozza:2002zj,Perlick:2004tq,LENS2,Bozza_2010,Bartelmann:2016}.

\section{Light scattering in the large distance approximation}

\n{9}

\subsection{Scattering angle}

In our previous considerations, we adopted the weak field approximations and considered $M/\ell$ as a small parameter, where $\ell$ is the impact parameter. However, we did not assume that the size of the body, $L$, is also small. When $L/\ell\ll 1$, the expression for the scattering angles greatly simplifies. Let us discuss this approximation.

Let us note that
\ba
&|z-z'|^2=|z|^2-2 z\cdot z'+|z'|^2\\
&=|z|^2\left( 1-2{ z\cdot z'\over |z|^2}+{|z'|^2\over |z|^2}\right)\, .
\ea
We assume that $|z|\sim \ell$ and $|z'|<L$. Then in the approximation $L\ll\ell$, one has
\be\n{LOG}
\ln|z-z'|^2=\ln |z|^2-2{ z\cdot z'\over |z|^2}+O((L/\ell)^2)\, .
\ee
Calculations give
\ba
&{\pa \over \pa z^a}\ln|z-z'|^2=2{z_a\over |z|^2}-{2\over |z|^4}q_{ab} z'^b\, ,\\
&q_{ab}=|z|^2\delta_{ab}-2z_a z_b\, .
\ea

Let us denote $d^3 V=dX'd^2z' $. Then the following relations are valid
\ba
&\int d^3V \rho=M\hh \int d^3V \pa_{a'} j^{a'}=0\, ,\\
&\int d^3V z'_b  \pa_{a'} j^{a'}=-\int d^3V j_{b'}=-J_{Xb}\, .
\ea
Using these results one finds the following expression for the scattering angle (\ref{INT})
\be
\Delta\theta_a=4{Mz_a\over |z|^2}\,+\,4q_{ab}{J_{Xb}\over |z|^4}\, .
\ee

One can always choose coordinates $z^a$ so that the 2D vector $J_{Xb}$ takes the form $J_{Xb}=J\delta_b^2$. In this case, the vector of the angular momentum $L^a$ is directed along the $z^3$ axis, $L^a=J\delta^a_3$.

For scattering in the "equatorial plane" when $z^3=0$ one obtains
\be\n{DDDD}
\Delta\phi\equiv \pi+ \Delta\theta_2=\pi+ {4GM\over \ell}-{4GJ\over \ell^2}\, .
\ee
Here we denote by $\ell$ the impact factor for the ray, so that $|z|^2=\ell^2$. We also restored the Newton's coupling constant $G$. This result is in complete agreement with the expression (\ref{MMJJS}) for null ray scattering in the equatorial plane in Kerr spacetime\footnote{
Let us note that in the linearized approximation which is used in the paper, the additional term
$\sim (GM/\ell)^2$  in (\ref{MMJJS}) proportional to $G^2$ is not present in (\ref{DDDD}). For the Kerr black hole this term is of the same order as $GJ/\ell^2$. However, the effect of the rotation can be distinguished from others by measuring the difference of the scattering angles for two values of the angular momentum $J$, or for two opposite values of the impact factor $\ell$ and $-\ell$ with the same value for $J$. For extended objects with size $L\gg GM$, the term proportional to $J$ can be larger than the $\sim(GM/\ell)^2$ contribution.}.

\subsection{Shift effect}

Let us now discuss an effect regarding the shift of the null rays associated with the components $A_{a}$ of the metric. We again use the large distance approximation. Substituting the expansion (\ref{LOG}) into (\ref{FFAAEQ}), one obtains
\be \n{SHIF}
{\cal A}_a=-2G e_{ab}\int d^3V \left(\ln |z|^2-2{ z\cdot z'\over |z|^2}\right) j^{,b'} \, .
\ee
Since
\be
\int d^3V \pa_{b'}j=0\, ,
\ee
only the second term in parentheses in expression (\ref{SHIF}) gives a non-vanishing contribution. After integrating this term by parts and restoring the Newton's constant $G$ one gets
\be
\Delta z_a={4GJ e_{ab}z^b\over |z|^2}\, .
\ee
This shift vector $\Delta z_a$ is orthogonal to $z^a$, and its norm is
\be
|\Delta z_a|={4GJ \over b}\, .
\ee
Here $b$ is the impact parameter, $|z|^2=b^2$.
This result correctly reproduces the expression (\ref{SSHI}) for the shift of azimuthal null rays in the Kerr metric.

\section{Discussion}

\n{10}

In this paper, we studied the gravitational field of an ultrarelativistic massive spinning object. The corresponding metric of such an object was obtained by taking the Penrose limit of the object's field in the weak field approximation. We demonstrated that for spinning matter, it is not sufficient to keep the energy density fixed.  One also needs to perform a proper rescaling of the angular momentum density components in the directions orthogonal to the direction of motion. We studied the scattering of free particles and light rays by the gravitational fields of the ultrarelativistic spinning objects, and constructed the map $\Psi$ relating the parameters of inward and outward trajectories in terms of the matter and spin distributions . We briefly discussed how the obtained results are related to the gravitational memory effect. We also showed that the constructed map $\Psi$ can be used to calculate scattering angles for light propagating in the vicinity of a massive spinning object in its own rest frame. The latter result allows one to recover the standard results for gravitational lensing, and obtain their generalization in the case where the matter producing this effect can have non-vanishing local differential rotation.

\appendix

\section{Null ray scattering in the Kerr geometry: weak field approximation}
\n{A}

\subsection{Scattering in the equatorial plane}

The geodesic equations in the Kerr metric have four integrals of motion. As a result, these equations are completely integrable and their solutions can be written in quadratures (see e.g. \cite{Misner:1974qy}). For null ray propagation in the equatorial plane, the solutions to the equations of motion can be written explicitly in the form of the elliptic integral \cite{BL_KERR}. When the impact parameter $\ell$ for the null ray is much larger than the gravitational radius $GM$ of the Kerr black hole, one can decompose the expression for the ray's scattering angle in powers of the small parameter $GM/\ell$. Here we reproduce this result by a different method, which we also use in the next subsection for azimuthal null rays.

The Kerr metric in the equatorial plane $\theta=\pi/2$ takes the form
\ba
& ds^2=g_{tt}dt^2+2g_{t\phi}dt d\phi+g_{\phi\phi}d\phi^2+ g_{rr} dr^2\, ,\\
&g_{tt}=-(1-{2M\over r}) \hh g_{t\phi}=-{2Ma\over r} \, ,\\
&g_{\phi\phi}=r^2+a^2+{2a^2M\over r}\hh g_{rr}=(1-{2M\over r}+{a^2\over r^2})^{-1}\, .
\ea
Here $M$ is the mass of the Kerr black hole, and $a$ is the rotation parameter related to the angular momentum of the black hole $J$ by $a=J/M$. In what follows, we shall use the dimensionless form of this parameter. Namely we denote by $s$ the following ratio
\be \n{RAP}
s={J\over M^2}\, .
\ee
We call $s$ the rapidity. For a black hole, $s$ takes values between 0 and 1.

Let $x^{\mu}=x^{\mu}(\tau)$ be a null geodesic in the equatorial plane, and let $\tau$ be its affine parameter. We denote by a dot the derivative with respect to $\tau$.
This 3D metric has 2 Killing vectors, $\pa_t$ and $\pa_{\phi}$, and their corresponding integrals of motion are
\ba
&E=-g_{tt}\dot{t}-g_{t\phi}\dot{\phi}\, ,\\
&L=g_{t\phi}\dot{t}+g_{\phi\phi}\dot{\phi}\, .
\ea
The third integral of motion is
\be
g_{\mu\nu}\dot{x}^{\mu}\dot{x}^{\nu}=0\, .
\ee
These three integrals of motion are sufficient for the complete integrability of the equations of motion in the equatorial plane. One can obtain the following equation for the null ray trajectory (for details see e.g \cite{FrolovZelnikov:2011})
\ba
&{d\phi\over dr}=-Q\hh
Q={((r-2M)L+2aME)\sqrt{r}\over \sqrt{P}\Delta}\, ,\\
&P=E^2(r^3+a^2r+2a^2 M)-4aMEL -(r-2M)L^2\, ,\\
&\Delta=r^2-2Mr+a^2\, .
\ea
 The ray trajectory has a radial turning point where $P(r)=0$. We denote a solution of this equation by $r_0$. Before the turning point, the sign in the expression for $d\phi/dr$ is negative. For the incoming ray at large $r$, one has
\be
{d\phi\over dr}\approx -{\ell\over r^2}\hh \ell=L/E\, .
\ee
Integrating this equation, one gets
\be \n{ppll}
\phi\approx \phi_0+\ell/r\, .
\ee
We choose the coordinate $\phi$ so that for incoming rays $\phi_0=0$. Equation (\ref{ppll}) shows that $\ell$ is the impact parameter. We assume that this parameter can be both positive and negative.

Let us denote by $P_0$ the value of function $P(r)$ at $r=r_0$
\be
P_0=E^2(r_0^3+a^2r_0+2a^2 M)-4aMEL -(r_0-2M)L^2\, .
\ee
By definition of the turning point, $P_0=0$. Hence,
\be
P\equiv P-P_0=(r-r_0)[(r^2+a^2+r r_0+r_0^2)E^2-L^2]\, .
\ee

It is convenient to use the following dimensionless variables $(y,m,s,\lambda)$ to rewrite $Q$ as a dimensionless quantity $\chi=Q r_{0}$
\be
r=r_0 y\, ,  M=mr_0\, , a=s M\, ,  \ell=\lambda r_0\, .
\ee
In these variables, we now have
\be\n{QQQ}
\chi={\sqrt{y}[ 2(s m-\lambda)m+\lambda y] \over \sqrt{y-1}
\sqrt{1+y+y^2+s^2 m^2-\lambda^2}
(y^2+s^2 m^2-2my)}\, ,
\ee
The equation for the turning point $P_0=0$ can be solved for in terms of $\lambda$, and one gets
\be \n{aml}
\lambda={\sqrt{1-2m+s^2 m^2}-2s m^2\over 1-2m}\, .
\ee

The dimensionless parameter $m$ is the mass of the black hole as measured in units $r_0$. However, $r_0$ is defined implicitly as a solution of the algebraic equation $P(r)=0$. It is convenient to rewrite this parameter in terms of a new dimensionless parameter $\mu=M/\ell$, which is defined in terms of the scattering parameter (impact factor) $\ell$. For this purpose, we use the following relation
\be\n{mmmm}
m=\mu \lambda\, .
\ee
As a result, the relation (\ref{aml}) takes the form of an equation relating two parameters, $\lambda$ and $\mu$
\be \n{aml1}
\lambda={\sqrt{1-2\mu\lambda +s^2 \mu^2 \lambda^2}-2s \mu^2 \lambda^2\over 1-2\mu \lambda}\, .
\ee
For small $\mu$, this equation can be solved perturbatively using the expansion
\be \n{lam}
\lambda=\sum_{k=0}^{\infty} {\mu^k \over k!}\lambda_k\, .
\ee
For $\mu=0$ one has $\lambda=1$, so $\lambda_0=1$.

Substituting (\ref{lam}) into (\ref{aml1}) allows one to find the coefficients $\lambda_k$. We present here the first several terms of the series (\ref{lam})
\ba\n{llll}
&\lambda=1+\mu+{1\over 2}(s^2-4s+5)\mu^2\\
&+(3s^2-10s+8)\mu^3+O(\mu^4)\, .
\ea

Substituting (\ref{llll}) and (\ref{mmmm})  into the function $\chi$, (\ref{QQQ}), and expanding the obtained expression into powers of $\mu$, one finds the coefficients $ \chi_k(y,s)$ of the following expansion
\be
\chi=\sum_{k=0}^{\infty} {\mu^k\over k!} \chi_k(y,s)\, .
\ee
The scattering angle can be found as follows
\be
\phi_s=2\int_{y=1}^{\infty} \chi\; dy\, .
\ee
Then one gets
\ba\n{mmss}
&\phi_s = \pi + 4\mu + \big(\frac{15}{4}\pi-4s\big)\mu^2 \\
&+ \big(\frac{128}{3}-10\pi s+4s^2\big)\mu^3+O(\mu^4)\, .
\ea

Using the definition of $\mu$ (\ref{mmmm}), and restoring the Newton's coupling constant $G$, one can write the first few terms of (\ref{mmss}) in the form
\be \n{MMJJS}
\phi_s = \pi + 4{GM\over \ell} -4{GJ\over \ell^2}+{15\pi\over 4}\left( {GM\over\ell}\right)^2+\ldots \, .
\ee

As we noted before, $M$ is the mass of the Kerr black hole, and $s$ is its rapidity parameter, which can take values between 0 and 1 and is given by (\ref{RAP}).

The first terms of this expression coincide with the result presented in \cite{BL_KERR}\footnote{Let us note that the sign of the rotation parameter in the paper \cite{BL_KERR} is chosen to be opposite to the sign adopted in \cite{Misner:1974qy,FrolovZelnikov:2011}.} See also \cite{Skrotskii_2,PLEBANSKI_KERR}.

\subsection{Scattering of azimuthal rays}

Let us now discuss the case where an incoming null ray moves parallel to the axis of symmetry of a Kerr black hole. We denote by $b$ its distance from this axis at a very large radius. We choose the angle $\phi$ so that for this incoming ray, $\phi=0$. We call $b$ the impact parameter.

To find the shift in the angle $\phi$ after the scattering, we proceed in the same way as in the previous subsection.
The equations of motion written in the first order form imply that
\ba \n{THET}
&{d\theta\over dr}=\pm {\sqrt{\Theta}\over \sqrt{R}}\, ,\\
&\Theta=q+\cos^2\theta(a^2-\ell^2/\sin^2\theta)\, ,\\
&R=(r^2+a^2-a\ell)^2-(q+(a-\ell)^2)\Delta\, ,\\
&\Delta=r^2-2Mr+a^2\, .
\ea
Here
\be
q={{\cal Q}\over E^2}\hh \ell={L\over E}\, .
\ee
Here $L$ is the angular momentum and ${\cal Q}$ is the Carter's constant.
For the incoming asymptotically azimuthal null ray at $r\to \infty$, $\theta\to \pi$. This is possible only when the angular momentum $\ell$ for such a ray vanishes. Thus, one has
\be
\Theta=q+a^2\cos^2\theta\hh
R=(r^2+a^2)^2-(q+a^2)\Delta\,
\ee

For the incoming ray at large $r$, one has $\theta\approx \pi -\Delta \theta$, and the relations (\ref{THET}) give
\be
{d\Delta\theta\over dr}\approx -{\sqrt{q+a^2}\over r^2}\, .
\ee
By integrating this equation, one finds
\be
\Delta\theta\approx \sqrt{q+a^2}/r\, .
\ee
Since $b=\lim_{r\to\infty} (r\Delta\theta)$, one has
\be
b=\sqrt{q+a^2}\, .
\ee
This relation allows one to express dimensionless Carter's constant $q$ in terms of the impact parameter $b$
\be \n{qqbb}
q=b^2-a^2\, .
\ee
The conservation of the parameter $q$ guarantees that in the outgoing regime, that is when $\theta\to 0$ and $r\to\infty$, the outgoing impact parameter coincides with the incoming one, $b$. Using (\ref{qqbb}), one gets
\be \n{RRR}
R=(r^2+a^2)^2-b^2\Delta\, .
\ee

Let us consider now the equation for the angle $\phi$
\be \n{pprr}
{d\phi\over dr}=-{2Mar\over \Delta\sqrt{R}}\, .
\ee
At the turning point, where $r$ reaches its minimum value $r=r_0$, one has
\be \n{PP00}
R_0\equiv R(r=r_0)=0\, .
\ee
One also has
\be
R\equiv R-R_0=r^4-r_0^4 +(2a^2-b^2)(r^2-r_0^2) +2Mb^2(r-r_0)\, .
\ee
Let us introduce the following dimensionless parameters
\be
y={r\over r_0}\hhh s  m={a\over r_0}\hhh m={M\over r_0}\hhh \lambda={b\over r_0}\, .
\ee
Then the equation (\ref{pprr}) takes the form
\ba \n{FFFF}
&{d\phi\over dy}=F\, ,\\
&F=-{2s m^2 y\over \sqrt{y-1}(y^2-2my+s^2 m^2)}{1\over \sqrt{N}}\, ,\\
&N=y^3+y^2+y+1 \\
&+(2s^2 m^2-\lambda^2)(y+1)+2m\lambda^2\, .
\ea
As we did earlier, we denote
\be \n{MOB}
\mu={M\over b}\, ,
\ee
so that $m=\mu\lambda$.
Then the equation  (\ref{PP00}) for the turning point can be solved to obtain the equation for $\lambda$
\be \n{LALA}
\lambda={1+s^2\mu^2\lambda^2\over \sqrt{1-2\mu\lambda+s^2\mu^2\lambda^2}}\, .
\ee

We assume that $b$ is sufficiently large, so the parameter $\mu$ allows one to keep track of the order of smallness of different terms. Namely, a term which contains $\mu^n$ is proportional to $b^{-n}$.  We use the following expansion for $\lambda$
\be \n{LLLL}
\lambda=\sum_{k=0}^{\infty} {\mu^k \over k!}\lambda_k\hh \lambda_0=1\, .
\ee
Then by solving equation (\ref{LALA}), one can find coefficients $\lambda_k$ which enter the expansion (\ref{LLLL}). For the first 4 coefficients, one gets
\ba
&\lambda_1=1\hh \lambda_2=5+s^2\hh \lambda_3=48+6s^2\, ,\\
&\lambda_4=693+42s^2+9s^4\, .
\ea

We then substitute (\ref{MOB}) and  (\ref{LLLL}) into the function $F$ (\ref{FFFF}), and expand the obtained expression into powers of $\mu$.  With this, one finds the coefficients $F_k(y,s)$ of the following expansion
\be
F=\sum_{k=0}^{\infty} {\mu^k\over k!} F_k(y,s)\, .
\ee
The total angular shift, that is the value of $\phi$ at $r=\infty$ and $\theta=0$, is
\be
\Delta\phi=2\int_{y=1}^{\infty} F dy\, .
\ee
Using an expansion of $F$ in powers of $\mu$, and calculating the obtained integrals, one gets
\be
\Delta\phi=\mu^2s\big(4+5\pi\mu+64\mu^2\big)+O(\mu^5)\, .
\ee
Hence, the leading term of $\Delta\phi$ is
\be \n{FINP}
\Delta\phi={4GJ\over b^2}\, .
\ee

Thus, as a result of scattering by a Kerr black hole with angular momentum $J$, the incoming azimuthal null ray with impact parameter $b$ acquires an angular shift in $\phi$ (\ref{FINP}), while the amplitude of the impact parameter for the outgoing ray remains the same as the incoming one. This angular shift is the result of frame-dragging of the ray induced by the rotation of the black hole. This result can be formulated in a slightly different way. Let $\vec{b}$ be a 2D impact parameter vector in the 2D plane orthogonal to the incoming ray. Then the impact vector for the outgoing ray is shifted in the direction of black hole rotation by the vector $\vec{\beta}$, which is orthogonal to $\vec{b}$. Its value is
\be\n{SSHI}
|\vec{\beta}|=\Delta\phi \ b={4GJ\over b}\, .
\ee

\section{Christoffel symbols and null geodesics}
\n{B}

Let us consider a metric of the following form
\be\n{gyr}
ds^2=-2 du\, dv+ dz_a dz^a+\Phi du^2+2A_a du \ dz^a\, .
\ee
We also remember that indices $a,b,\ldots$ take values 2 and 3. The metric in the 2D space spanned by these coordinates is $\delta_{ab}$, so that $z^a=z_a$ and $A_a=A^{a}$.

We assume that metric coefficients $\Phi$ and $A_a$  have the following form
\be\n{PPAA}
\Phi=p(u)F(z)\hh A_a=q(u) {\cal A}_a(z)\, .
\ee
Then the exact non-vanishing Christoffel symbols calculated for this metric (\ref{gyr}) are
\ba\n{GGG}
&\GG{u}{u}{v}=-{1\over 2}\dot{p}F+q\dot{q}{\cal A}_{a}{\cal A}^{a}-{1\over 2}qp{\cal A}_{a}F^{,a}\, ,\\
&\GG{u}{u}{a}=\dot{q}{\cal A}^{a}-{1\over 2}pF^{,a}\, ,\\
&\GG{u}{a}{v}=-{1\over 2}pF_{,a}-{1\over 2}q^{2}{\cal F}{\cal A}_{b}e_{a}^{\ b}\, ,\\
&\GG{u}{a}{b}=-{1\over 2}q{\cal F}e_a^{\ b}\, ,\\
&\GG{a}{b}{v}=-{1\over 2}q({\cal A}_{a ,b}+{\cal A}_{b ,a})\, .
\ea
Here  $e_{ab}$ is a 2D Levi-Civita symbol with $e_{23}=1$, and
\be
{\cal F}=e^{ab}{\cal A}_{a ,b}\, .
\ee
We also denote by a dot the derivative with respect to $u$.

Let us consider a geodesic line $x^{\mu}(\tau)$ in the metric (\ref{gyr}). One has
\be \n{norm}
g_{\mu\nu}{dx^{\mu}\over d\tau}{dx^{\nu}\over d\tau}=-\varepsilon\, .
\ee
Here $\varepsilon=1$ for a massive particle and $\varepsilon=0$ for a null ray.
This normalization condition means that for a massive particle, the parameter $\tau$ coincides with proper time $\tau$. For a null ray, $\tau$ is an affine parameter.

The geodesic equations are
\be \n{GEOD}
{d^2 x^{\mu}\over d\tau^2}+\GG{\nu}{\lambda}{\mu} {dx^{\nu}\over d\tau}{dx^{\lambda}\over d\tau}=0\,  .
\ee
Since $\pa_v$ is a Killing vector of the metric (\ref{gyr}), one has
\be
{\cal U}\equiv {du\over d\tau}=\const \, .
\ee
For a null ray, using the ambiguity in choice of the affine parameter, one can put this constant ${\cal U}$ to be equal to 1, so that $\tau=u$. For a massive particle, ${\cal U}$ is an additional independent parameter.

It is convenient in both cases, for massive and massless particles, to use the parameter $u$ instead of $\tau$. In this parametrization, one writes $x^{\mu}=x^{\mu}(u)$, and the normalization condition (\ref{norm}) takes the form
\be \n{VVV}
2\dot{v}=\dot{z}_a \dot{z}^a+pF+2q{\cal A}^a\dot{z}_a+{\varepsilon\over {\cal U}^2}\, .
\ee
Using expressions for the Christoffel symbols (\ref{GGG}), one can write 2 equations (\ref{GEOD}) for $\mu=a$ as follows
\be \n{ZZZ}
\ddot{z}_a+q{\cal F} e_{ab}\dot{z}^b+\dot{q}{\cal A}_a-{1\over 2}pF_{,a}=0\, .
\ee
Let us note that this equation does not contain $v$ or $\dot{v}$. Hence, this is a consistent set of two second order ordinary differential equations. After solving these equations, one can find $v(u)$ by solving equation (\ref{VVV}).

\section*{Acknowledgments}

The authors thank the Natural Sciences and Engineering Research Council of Canada and the Killam Trust for their financial support. We also are grateful to Andrei Zelnikov for useful discussions and his help with preparation of the figures.
\vfill


%

\end{document}